\documentclass[twocolumn]{aastex63}
\usepackage{graphicx}
\usepackage{txfonts}
\usepackage{setspace} 
\usepackage[inline, shortlabels]{enumitem}

\def\be{\begin{equation}}
\def\bea{\begin{eqnarray}}
\def\eea{\end{eqnarray}}
\def\ee{\end{equation}}

\received{}
\revised{}
\accepted{}
\submitjournal{ApJL}

\shorttitle{GRB Prompt spectra in Backscattering model}
\shortauthors{Vyas et al.}
\graphicspath{{./}{figures/}}

\begin{document}
\setstretch{1.0}
\title{Predicting spectral parameters in the backscattering dominated model for the prompt phase of GRBs}

\correspondingauthor{Mukesh Kumar Vyas}
\email{mukeshkvys@gmail.com}




\author{Mukesh K. Vyas$^1$, Asaf Pe'er}
\affiliation{Bar Ilan University, \\
Ramat Gan, \\
Israel, 5290002 }
\author{David Eichler}
\affiliation{Ben-Gurion University, \\
Be'er Sheva, Israel, 84105}



\begin{abstract}
We present new results of the backscattering dominated prompt emission model in which the photons generated through pair annihilation at the centre of a gamma ray burst (GRB) are backscattered through Compton scattering by an outflowing stellar cork.
Using Comptonized pair annihilation spectrum accompanied by bremsstrahlung radiation for seed photons, we show that the obtained spectra produce low energy photon index \textbf{in the range }$\alpha \sim -1.95$ to $-1.1$, steeper high energy slopes $\beta \sim -3.5$ to $-2.4$ and spectral peak energies $\sim$ \textbf{few KeV to few $10\times$ MeV}. These \textbf{findings} are consistent with the values \textbf{covered} in GRB prompt phase observations.
\end{abstract}


\section{Introduction}
In the current understanding of long gamma ray bursts (GRBs), a collapsing core of a massive star (e.g., a Wolf Rayet star) leads to the observed GRB phenomenon \citep{1993ApJ...418..386L, 1993AAS...182.5505W, 1999ApJ...524..262M} while two compact stars gravitationally merge to produce a short GRB. In both the cases, a double sided jet is produced from the centre of the burst. As it propagates through the envelope of the collapsing star, this jet collects material ahead of it thereby forming a dense stellar cork which expands ahead of it. This cork is less energetic in short GRBs compared to long GRBs \citep{2017ApJ...834...28N}. After crossing the stellar envelope, the jet eventually pierces through the cork and escapes the system \citep{2002MNRAS.331..197R, 2003ApJ...586..356Z, 2004ApJ...608..365Z, 2014ApJ...784L..28N}. Energetic electrons inside the jet produce the observed signal responsible for the GRB prompt phase.
The observed spectrum is often interpreted in the framework of synchrotron radiation \citep{1993ApJ...415..181M,1996ApJ...466..768T,1998ApJ...494L.167P, 2015AdAst2015E..22P,2015PhR...561....1K}. 

An alternate picture of the prompt phase was proposed and developed by \cite{2008ApJ...689L..85E, 2014ApJ...787L..32E, 2018ApJ...869L...4E, 2021ApJ...908....9V} according to which most of the photons are produced at the centre of the star near the time of the burst through pair annihilation, in a plasma dominated by $e^\pm$ pairs. Pair annihilation in this plasma naturally produces a radiation pattern having an equilibrium temperature around a few MeV \citep{1986ApJ...308L..47G, 1986ApJ...308L..43P,2014ApJ...787L..32E,2018ApJ...869L...4E}. The cork in this picture, after being pushed by the expanding gas and radiation pressure, moves with relativistic speed ahead of the radiation beam emitted by the pair plasma. The radiation beam is not able to pierce through it, and most of the photons are reflected backward. Due to the relativistic aberration, these photons are beamed towards the motion of the cork before being detected by the observer.

{In previous attempts of incorporating comptonization in the GRB atmosphere, \cite{1994ApJ...428...21B} assumed a power law spectrum as seed photons' distribution and studied its attenuation through the medium assumed above the burst. He explained spectral features of the burst including the spectral peak energies.  
\cite{2011A&A...526A.110D} assumed seed photons having synchrotron spectrum and studied its modification due to Compton scattering in the burst atmosphere. 
Compared to these works, here we do not consider synchrotron or nonthermal power law process. Rather, our setup assumes a thermal (Maxwellian) distribution of pairs which emit photons via annihilation in the inner region of an empty jet funnel. Annihilation spectrum intrinsically has bremsstrahlung contribution and the photons are Comptonized within the pair plasma to produce the final seed spectrum. These photons then propagate through the jet funnel, after which they interact with the outflowing cork, producing the observed prompt GRB signal. As we show below, the obtained spectrum has  a negative low energy photon index. Further, power laws at high energy are generated due to multiple Compton scattering of the seed photons inside the cork.
In \cite{2021ApJ...908....9V} (Hereafter VPE21), we showed that multiple scattering of photons inside the cork that scattered from different angles with respect to the observer can explain some key observations such as Amati correlation [a correlation between spectral peak energies $\varepsilon_{peak}$ and equivalent isotropic energies $\varepsilon_{iso}$; see \cite{2006MNRAS.372..233A, 2021MNRAS.501.5723F}] and spectral lag, which could not be addressed in other works.}

However, following the assumption of monoenergetic seed photons, the obtained low energy slopes in VPE21 were {positive and hence} deviating compared to the observed slopes.
In this letter, we resolve this problem by considering {Comptonized} pair annihilation spectra for the seed photons at the centre of the burst {as explained above}. With this modification, the typical magnitudes of obtained low energy spectra are consistent with observations. 

In section \ref{sec_model} we briefly describe the model and proceed to detail of assumed  electron positron pair annihilation spectrum for seed radiation field in section \ref{sec_seed_dist}. We discuss the results in section \ref{sec_results} before summarizing the paper in section \ref{sec_summary}.

\begin {figure}[h]
\begin{center}
 \includegraphics[width=7cm, angle=0]{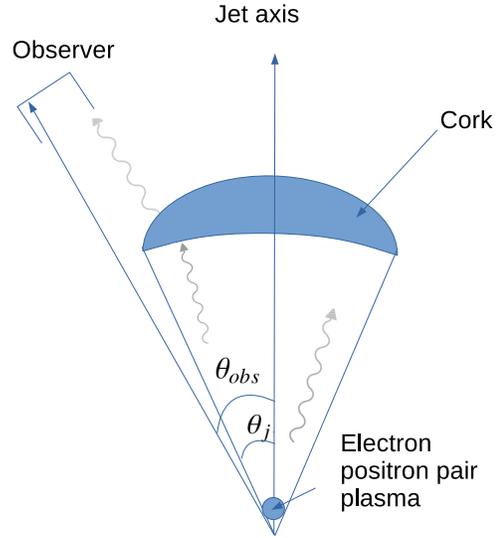}
\caption{Geometry of the system. Source of the radiation is electron positron pair plasma producing annihilation spectrum at the centre of the burst.The photons enter the cork that expands with Lorentz factor $\gamma$ and temperature $T_c$. The opening angle of the jet (and cork) is $\theta_j$ while the observer, situated at $\theta_{obs}$, observes photons that are scattered backwards by the inner surface of the cork.}
\label{lab_geom}
 \end{center}
\end{figure}
\section{Brief picture of the backscattering dominated model}
\label{sec_model}
The seed photons' produce near the centre of the burst due to pair annihilation and Comptonization. The detailed process of which is described in section \ref{sec_seed_dist}. 
These photons propagate inside the empty jet funnel to radially enter an optically thick cork with an opening angle $\theta_j$. The cork adiabatically expands with a constant Lorentz factor $\gamma$ and temperature $T_c$ at initial distance $r_i$ from the centre of the star (Figure \ref{lab_geom}). We carry out Monte Carlo simulations for studying the interaction of these photons with the relativistic electrons inside the cork. These photons may go through multiple Compton scattering with the energetic electrons before escaping through the cork's back surface. If the photons do not escape within $25$ scatterings, we consider them to be lost inside. Following the relativistic motion of the cork, all escaped photons are relativistically beamed in the forward direction and a fraction of these photons is observed by an observer situated at an angle $\theta_{obs}$ from the jet axis and at azimuth $\phi_{obs}$. The detected photons thus produce a spectral as well as a temporal evolution (light curve). Here, $\theta$ and $\phi$ are the spherical coordinates measured from the centre of the star. The system possesses azimuthal or $\phi$ symmetry.  We extend the work of  VPE21, considering relativistic {Comptonized }electron positron pair annihilation spectrum with temperature $T_r$ as a source of seed photons and reproduce the spectra.
Other details of the model are identical to those given in VPE21.
\section{Seed photon distribution : Comptonized Pair annihilation spectrum}
\label{sec_seed_dist}
{In a collapsing star, free neutrinos are generated and annihilate near the centre of the star in an empty funnel behind the outflowing jet. The neutrino annihilation near the centre of the star produces a copious amount of electron positron ($e^-e^+$) pairs \citep{1993ApJ...405..273W,1999ApJ...518..356P,1999ApJ...524..262M, 2003ApJ...594L..19L, 2005astro.ph..6368M, 2005astro.ph..6369M, 2014ApJ...796...26G}. The $e^-e^+$ pairs fall towards gravitating centre below a stagnation surface due to gravity and they escape outwards above it [see eg., Figure 1 of \cite{2005astro.ph..6368M}]. This plasma is hot with relativistic temperatures, and the pairs are in equilibrium with radiation produced within the plasma \citep{2013ApJ...770..159L}. The pair plasma produces a pair annihilation spectrum and associated bremsstrahlung radiation. This spectrum is further modified due to Compton scattering within the plasma. The emerging spectra that follow bremmstrahlung and Comptonization from thermal distribution of plasma at temperature $T_r$ were studied by \cite{1984PhST....7..124Z} through Monte Carlo simulations. There, he showed that the resultant spectrum at relativistic temperatures is flat in nature and decays exponentially at high frequencies.
For a typical plasma with density $n=2\times 10^{18}$ cm$^{-3}$, $\Theta_r=k_BT_r/m_ec^2$, and escape optical depth $\tau=1$.
We obtain the following numerical fit to the respective spectrum integrated over the emitting surface,
\be 
F_{\varepsilon}=C_0 \exp \left(-\frac{C_1 \varepsilon^2}{\Theta_r^2}\right) {~~~\rm KeV/s/}{\rm KeV}^{-1}
\label{eq_zd_fit}
\ee
Here $\varepsilon$ is energy of the photons normalized to electron's rest energy $m_ec^2$, $k_B$ is Boltzmann's constant, $m_e$ is the mass of the electron, $c$ is the light speed and $C_1=0.045$. $C_0(=2\times 10^{40}$ KeV s$^{-1}/$KeV) is a normalization parameter that depends on the pair density. As long as the plasma is relativistic it is independent of $T_r$. Note that its value does not affect the overall spectral shape, hence its exact parametric dependence will not affect the results presented here. 
Our Equation \ref{eq_zd_fit} provides good fit to the data presented by \cite{1984PhST....7..124Z} in his Figures 1(d) and 1(e). We find that this fit is applicable for relativistic temperatures $\Theta_r>0.3$ where the emergent spectra are flat.}
 
{These photons, then, propagate and enter the outflowing optically thick cork with temperature $\Theta_c=k_BT_c/m_ec^2$.
Further, the outcome of this seed spectrum intrinsically considers a constant temperature $\Theta_r$. It is a reasonable assumption as long as we are considering the prompt phase spectra where only initial temperature of the pair plasma is important. Thus we retain the assumption of delta function in injection time used in the previous paper. Later evolution of $\Theta_r$ to lower temperatures, related emission and their scattering with the cork may contribute to GRB emissions at late times $i.e.,$ afterglows and are beyond the current scope of this letter.}

\begin {figure}[h]
\begin{center}
  \includegraphics[width=7.5cm, angle=0]{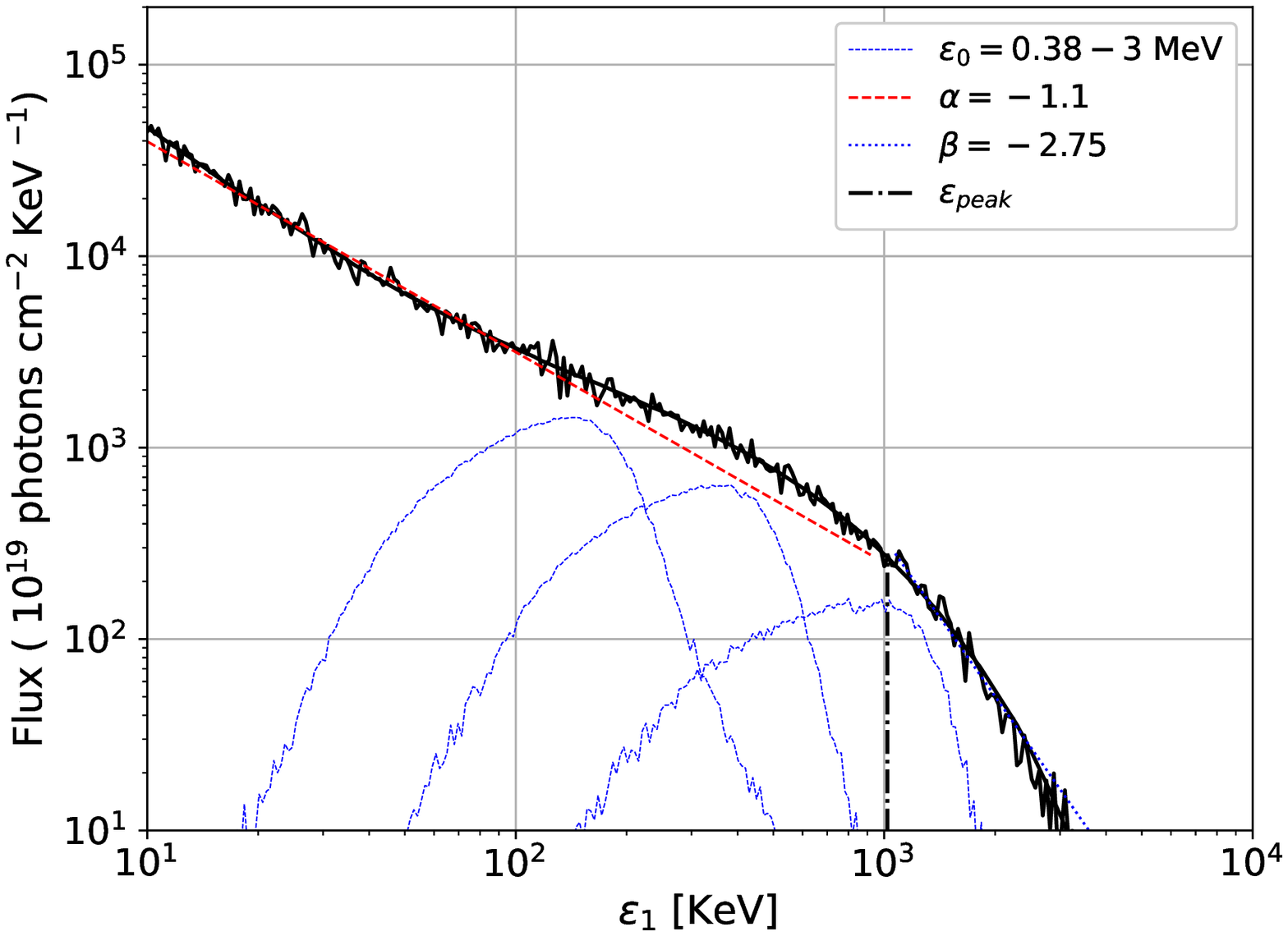}
  \includegraphics[width=7.5cm, angle=0]{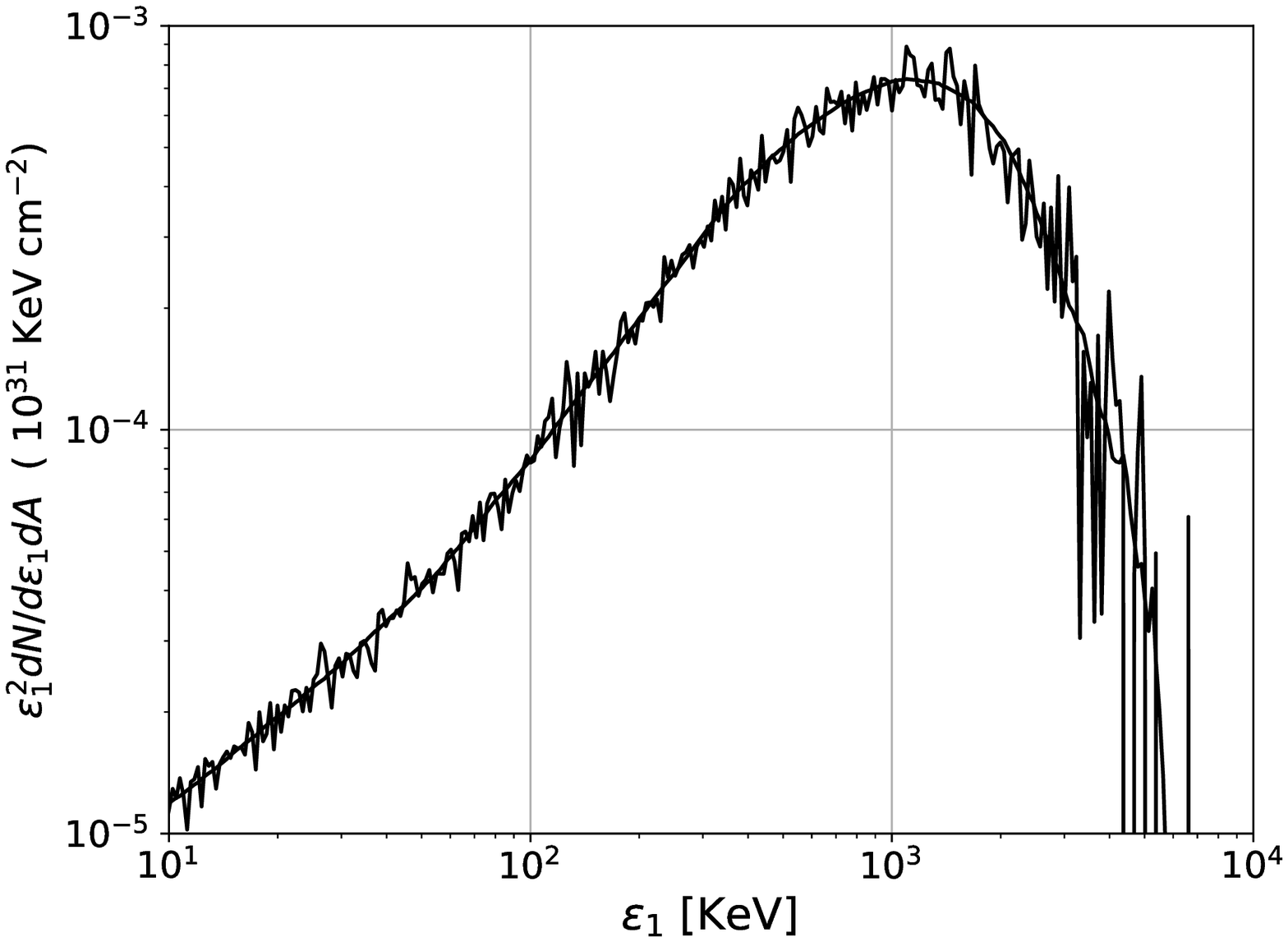}
\caption{Upper panel : Photon spectrum (solid black) for $\gamma=100$ and $\theta_{obs}=0.175$ obtained for pair temperature $\Theta_r=3$. \textbf{The jet opening angle is $\theta_j=0.1$ rad.} It is fitted for low energy photon index $\alpha=-1.1$, $\beta=-2.75$. Both the photon indices are connected at spectral peak energy $\varepsilon_{peak}=1020$ KeV. Blue dotted curves are corresponding monoenergetic photons for $\varepsilon_0=378,1000$ and $3000$ KeV. Lower panel : corresponding spectral emissivity $\epsilon_1^2 dN/d\epsilon_1$. \textbf{Overplotted solid black curve in both the panels is obtained by applying Savitzky–Golay filter for data smoothing}}.
\label{lab_Spectrum_th_0.05}
 \end{center}
\end{figure}

\section{Results}
\label{sec_results}

\begin {figure}[h]
\begin{center}
  \includegraphics[width=7.5cm, angle=0]{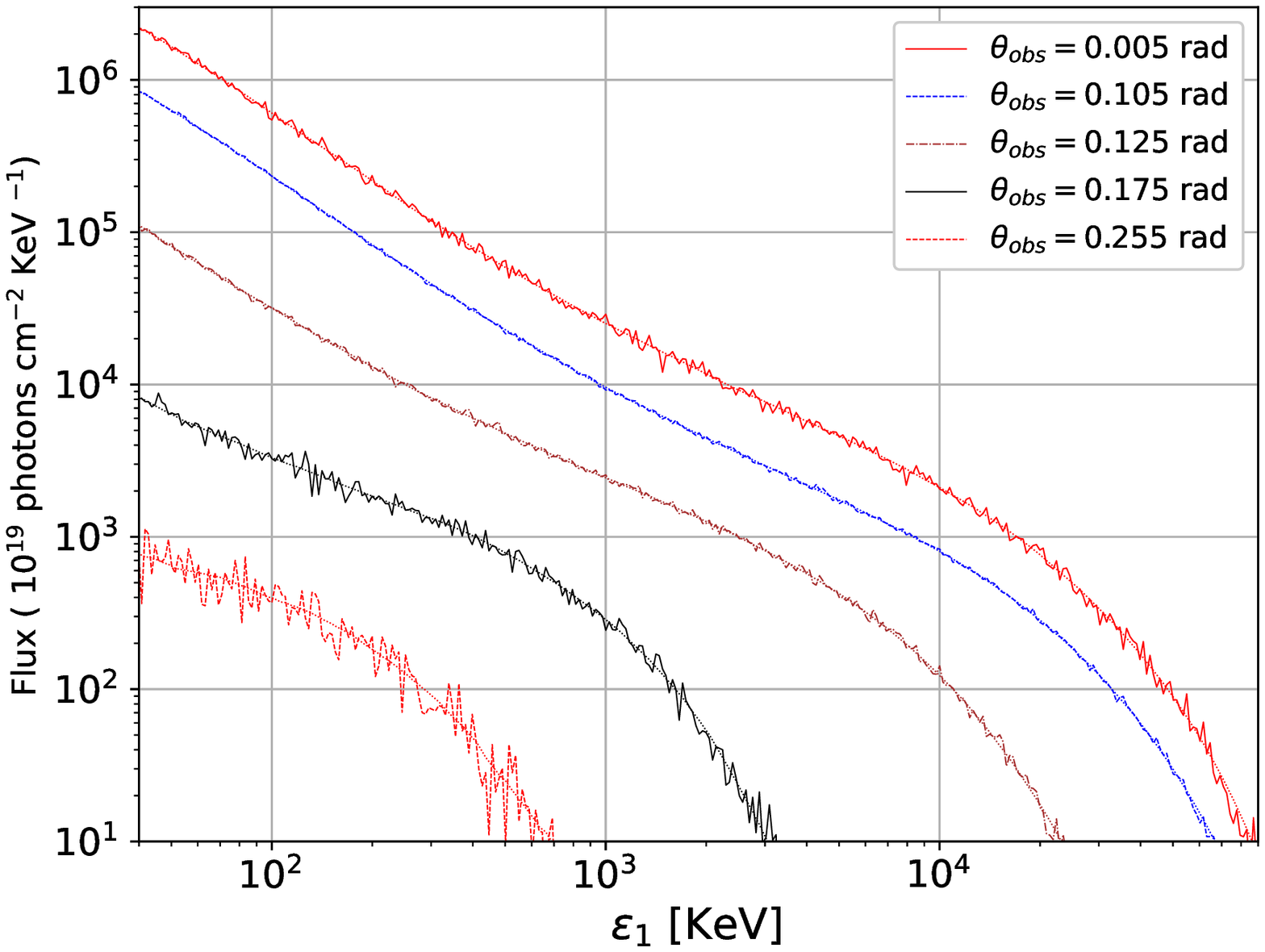}
 \includegraphics[width=7.5cm, angle=0]{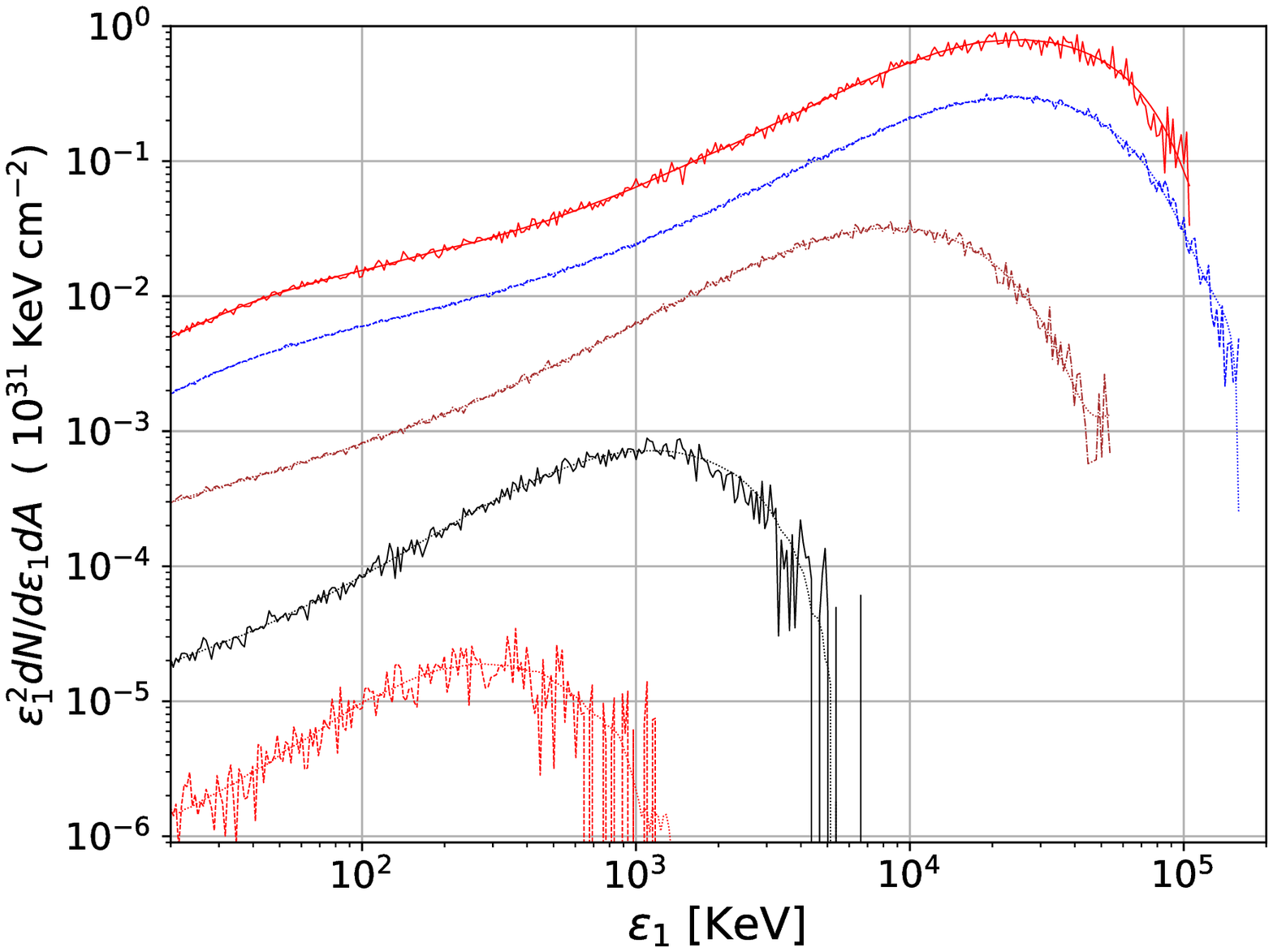}
  \includegraphics[width=7.5cm, angle=0]{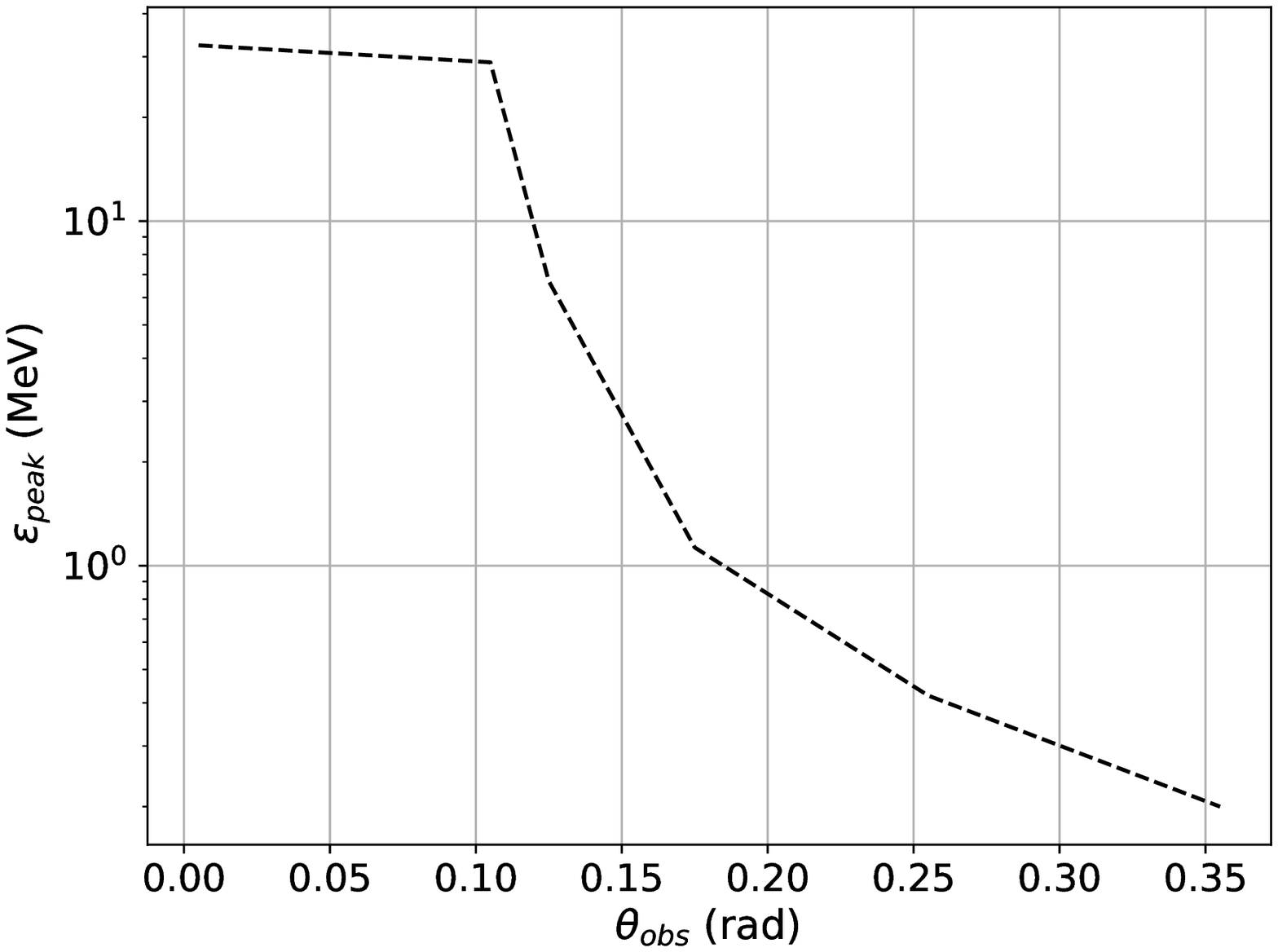}
\caption{\textbf{Variation of spectra (top and middle panels) with  observing angles in the range $\theta_{obs}=0.005-0.35$ rad. Overplotted dotted curves show the corresponding smoothed spectra by applying Savitzky–Golay filter. Variation of $\varepsilon_{peak}$ with $\theta_{obs}$ (bottom panels). Other parameters are same as Figure \ref{lab_Spectrum_th_0.05}.}}
\label{lab_Spectrum_Obs_vary_multi}
 \end{center}
\end{figure}

\begin {figure}[h]
\begin{center}
  \includegraphics[width=7.5cm, angle=0]{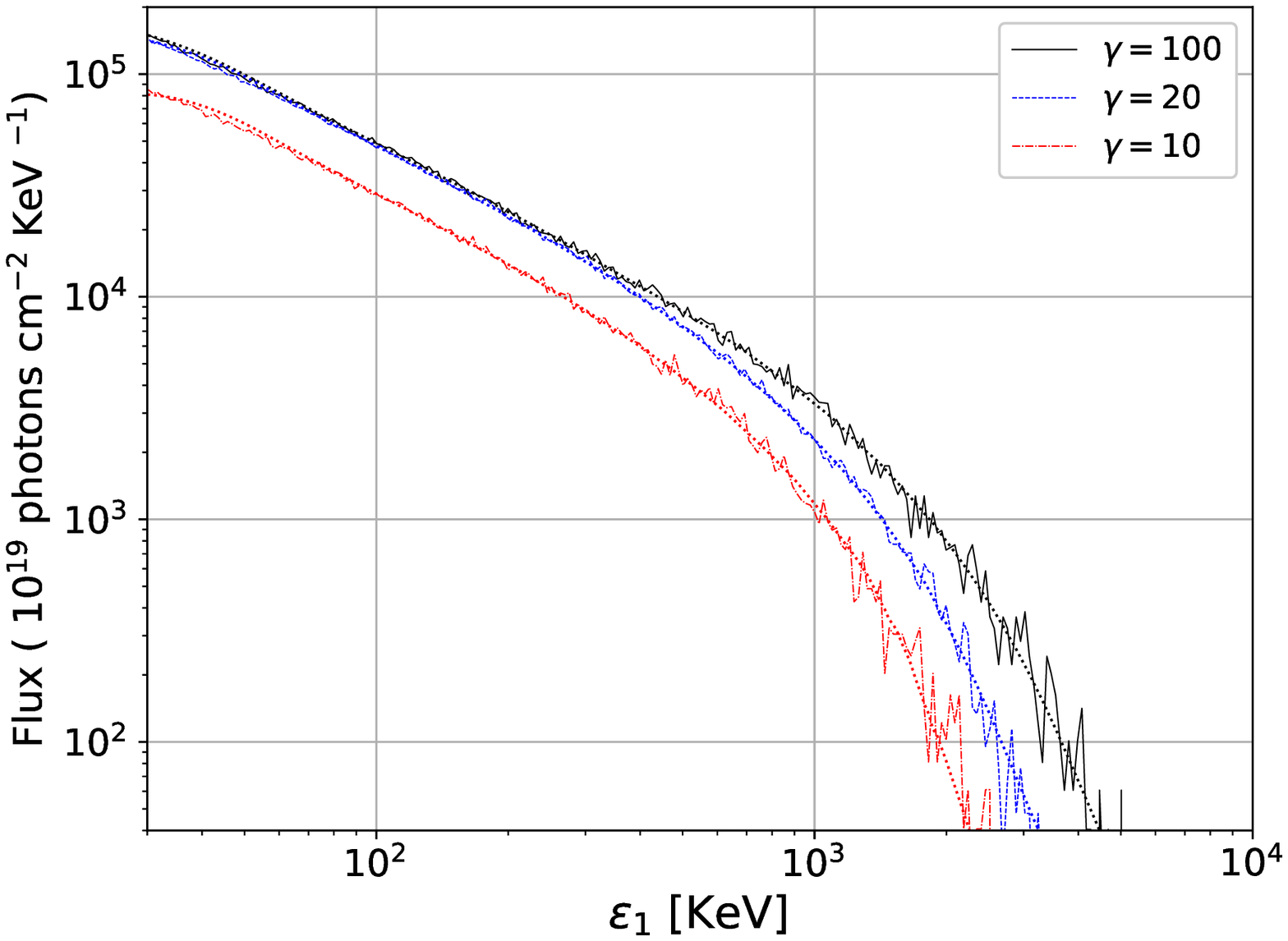}
  \includegraphics[width=7.5cm, angle=0]{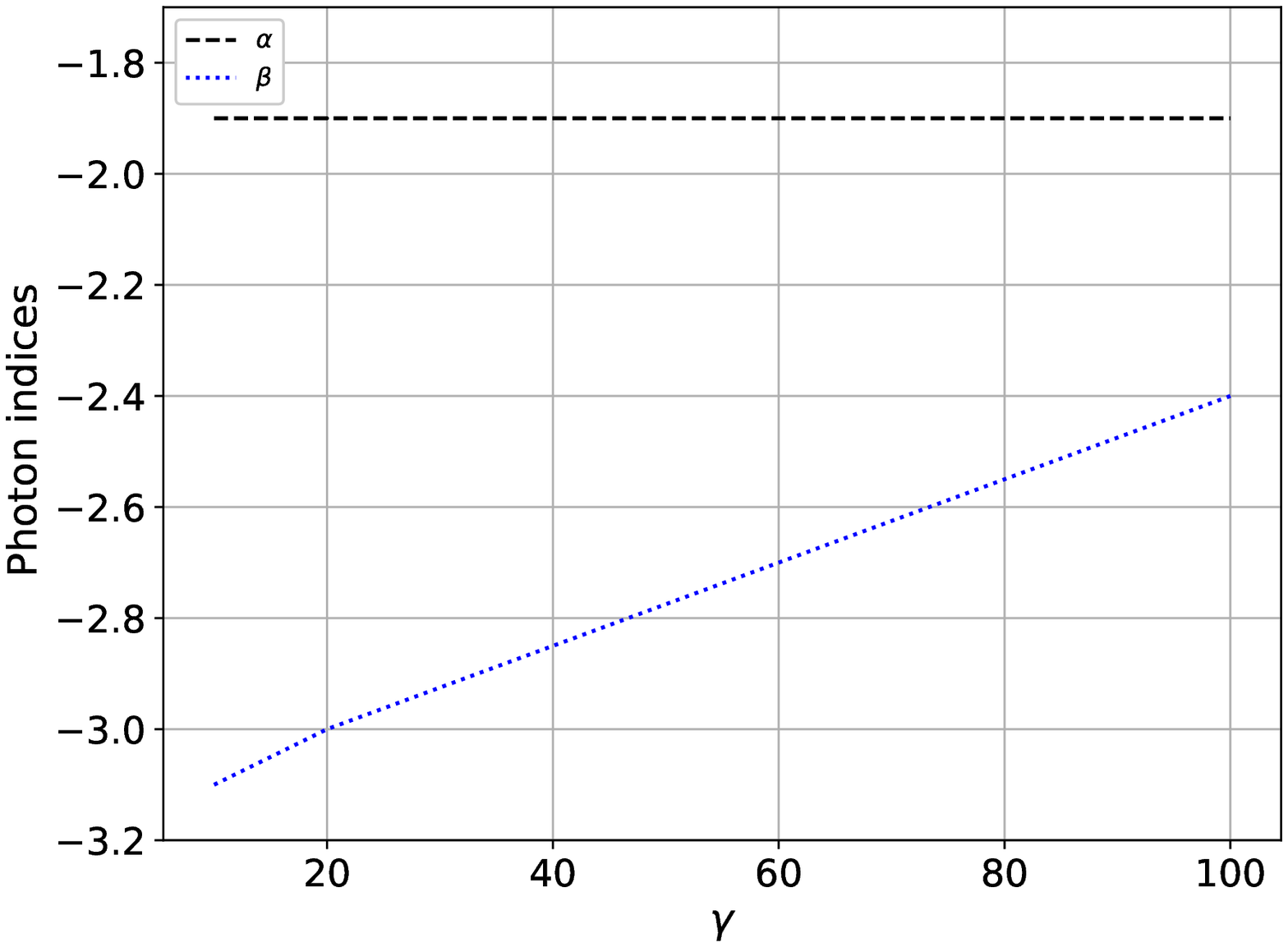}
\caption{\textbf{Top panel : Spectra for various Lorentz factors $\gamma=10,20,100$ for the pair temperature $\Theta_r=3$, cork temperature $\Theta_c=0.06$ and $\theta_{obs}=0.105$ rad. Overplotted dotted curves are obtained smoothed data points after applying Savitzky–Golay filter. Bottom panels : variation of $\alpha$ and $\beta$ with $\gamma$ for the same parameters. }}
\label{lab_Spectrum_gamma_vary_multi}
 \end{center}
\end{figure}

\subsection{General appearance of the spectrum}
\textbf{The cork forms effectively below the surface of the star with a typical radius $r_s = 10^{10}-3\times 10^{12}$cm. It is accelerated above it reaching an uncertain Lorentz factor which can be a few tens at a distance $r_i\ge r_s$ from the centre of the star.
Is was shown by \cite{2003ApJ...584..390W} that the cork may reach terminal Lorentz factor as high as 100 under specific conditions. However, other authors argue that a more typical value of the terminal Lorentz factor is only a few tens [see \cite{2003ApJ...586..356Z, lopez2013three} for details]. Hence, lacking a complete theory, in this manuscript, we explore the emerging spectra from a range of possible terminal Lorentz factors, $\gamma \sim 10-100$. We further consider the cork to expand radially and lose its energy adiabatically at a distance $~r_i=10^{12.5}$cm from the centre of the star.}
To generate \textbf{a typical} resultant spectrum, we consider a constant Lorentz factor of the cork $\gamma=100$, with opening angle $\theta_j=0.1$ rad and a temperature $\Theta_c=0.4$.\textbf{ 
Approximately $\sim$26 million photons are injected in the code and scaled with the burst having energy $10^{50}$ erg to calculate the fluxes. Hence the spectra are shown in the rest frame of the burst and are independent of cosmic redshift.}

In the upper panel of Figure \ref{lab_Spectrum_th_0.05}, the spectrum obtained for seed distribution of Comptonized pair annihilation and bremsstrahlung spectrum [according to equation \ref{eq_zd_fit}] is shown by black solid curve. To explain how this spectrum is generated, we plot the scattered spectra obtained for three different cases of monoenergetic photons, $\varepsilon_0=378,1000$ and $3000$ KeV (blue dotted curves); this is the setup considered in VPE21. Here the number of photons at each energy $\varepsilon_0$ are supplied according to the Comptonized pair annihilation spectral distribution of photons $dN/dt d\varepsilon=(F_{\varepsilon}/\varepsilon)$ in Equation \ref{eq_zd_fit}. It can be seen that the resulting spectrum is a superposition of the spectra generated by monoenergetic seed photons. The blueshifted peaks represent the increasing values of $\varepsilon_0$. In the lower panel, corresponding spectral emissivity $\varepsilon^2dN/d\varepsilon_1dA$ (KeV cm$^{-2}$) is plotted. 
\textbf{We used Savitzky–Golay filter to show the nature of the spectra by data smoothing. The filtered curves are shown by overplotted solid curves in both the spectra.}
For the parameters used, the spectral peak energy $\varepsilon_{peak}$ is obtained at $1020$ KeV and it separates the two spectral regimes with slopes $\alpha$ and $\beta$. All three monoenergetic spectra have positive low energy photon indices $\alpha=1$ while the resultant spectrum produces a negative slope $\alpha=-1.1$. The high energy photon index is obtained to be $\beta=-2.75$.  This result is similar to the observational result of the prompt GRB spectra that show $\alpha=-1$ and  $\beta=-2.5~$\citep{2006ApJS..166..298K, 2015AdAst2015E..22P}. In our model, generation of power laws at high energy follows from multiple scattering inside the cork; a complete explanation for this part of the spectrum appears in section 3.2.1, [eg. Figure 4] of VPE21. The obtained peak energies are also in accordance with the most abundant observed values for redshifted corrected spectra [see Figure 3 of \cite{2011A&A...531A..20G}].
\\
\subsection{Parametric dependence and explaining the observed spectra}
\textbf{The spectrum shown in Figure \ref{lab_Spectrum_th_0.05} is generated for specific parameters. Due to the high uncertainty in the theoretical models describing the formation,  acceleration and composition of the cork, there is a high uncertainty in  a number of key physical parameters describing the system. We here show that this uncertainty, by large, has only a moderate effect on the observed signal. In this respect, we analyze the dependence of
$\alpha, \beta$ and $\varepsilon_{peak}$ on system variables like cork Lorentz factor $\gamma$, cork temperature $\Theta_c$, the pair temperature $\Theta_r$ on which the seed photon's energy depends and the observer's angle $\theta_{obs}$.}
\subsubsection*{Dependence of observed spectra on observing angle $\theta_{obs}$}
\textbf{In the top two panels of Figure \ref{lab_Spectrum_Obs_vary_multi} we plot respective spectra (flux and spectral emissivity) as seen by different observers situated at different observing positions $\theta_{obs}(=0.005-0.355$ rad). All other parameters are kept identical to Figure \ref{lab_Spectrum_th_0.05}. In the bottom panel, we show the variation of $\varepsilon_{peak}$ with $\theta_{obs}$.
The observer's position  doesn't change the spectral slopes and hence $\alpha$ and $\beta$ are constant for all the observers situated at different angular positions. However, due to relativistic beaming, the observers at larger angular positions receive less flux as well as the spectral peak becomes softer. Variation of $\varepsilon_{peak}$ with $\theta_{obs}$ is shown in the last panel. For the observers within the jet angle $\theta_j=0.1$ rad, $\varepsilon_{peak}$ is roughly constant (few 10 MeV) while it monotonically decreases for $\theta_{obs}>\theta_j$  and falls upto few KeV for $\theta_{obs}\sim 0.35$ rad. }
\subsubsection*{Spectral evolution with cork Lorentz factors ($\gamma$)}
\textbf{In the top panel of Figure \ref{lab_Spectrum_gamma_vary_multi}, we consider three values of $\gamma=10,20$ and $100$ and plot the spectrum scattered by a cork with temperature $\Theta_c=0.06$. These spectra are seen by an observer at $\theta_{obs}=0.105$rad. The spectral shape is not very sensitive to $\gamma$ as the seed photons are first redshifted in the cork frame by a factor of $2\gamma$ and then after backscattering, these are again blueshifted for an on axis observer by the same amount. However, the flux received from less relativistic corks significantly decreases due to less effective relativistic beaming towards the observer. The low energy photon indices are unaffected by $\gamma$.
In the bottom panel, we show that the spectra are harder showing smaller magnitudes of $\beta$ for larger $\gamma$. This can be understood as the seed photons transforms to less energy for higher $\gamma$ and thus these photons are more efficiently inverse Comptonized to gain energy inside the cork thereby making the spectrum harder. Variation of $\gamma$ between 10-100 thus produces range of $\beta=-3.1$ to $-2.4$.}
\subsubsection*{Effect of plasma temperature $\Theta_r$ and cork temperatures $\Theta_c$ on the spectra}
\begin {figure*}[h]
\begin{center}
  \includegraphics[width=7cm, angle=0]{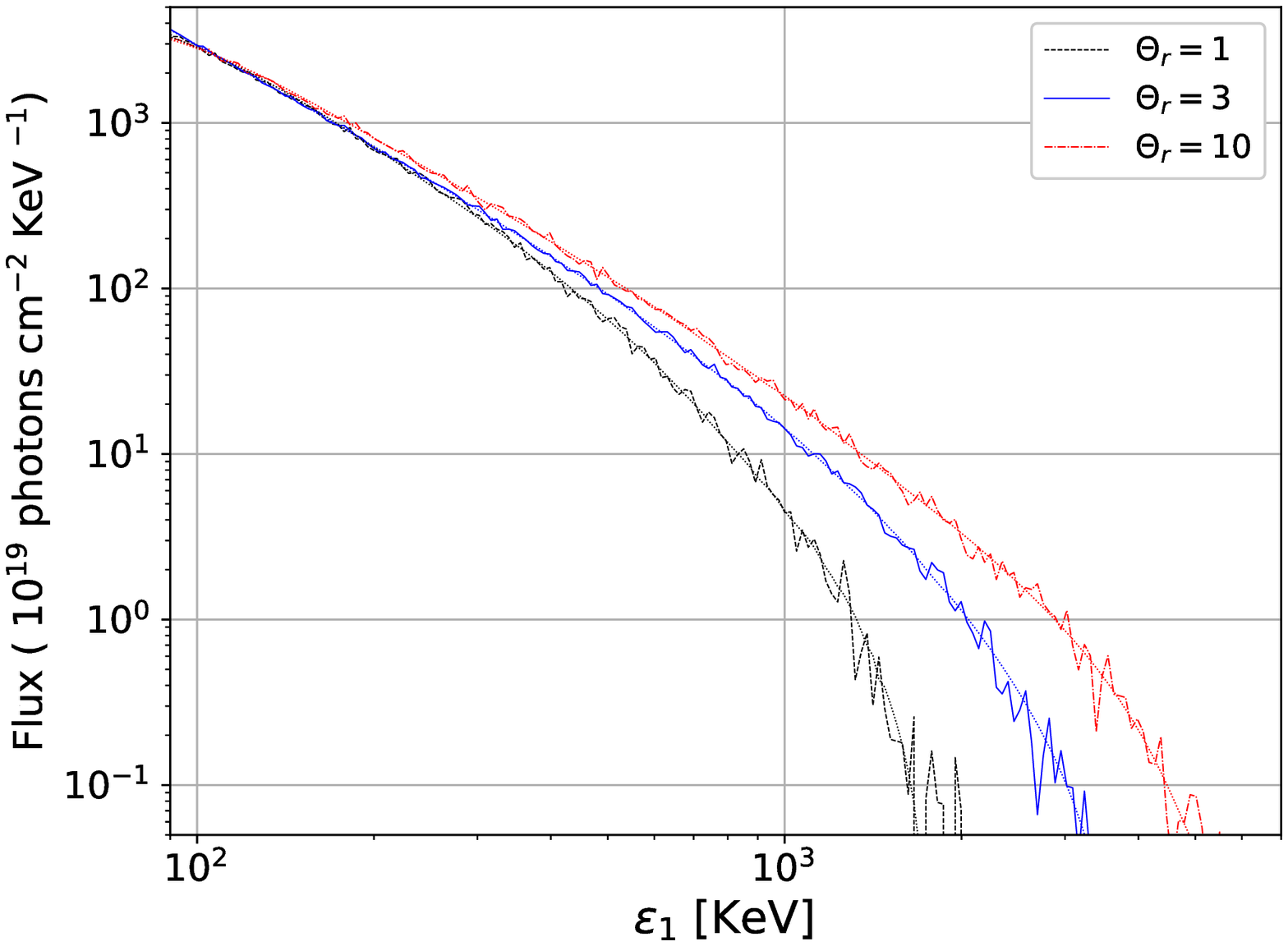}
    \includegraphics[width=7cm, angle=0]{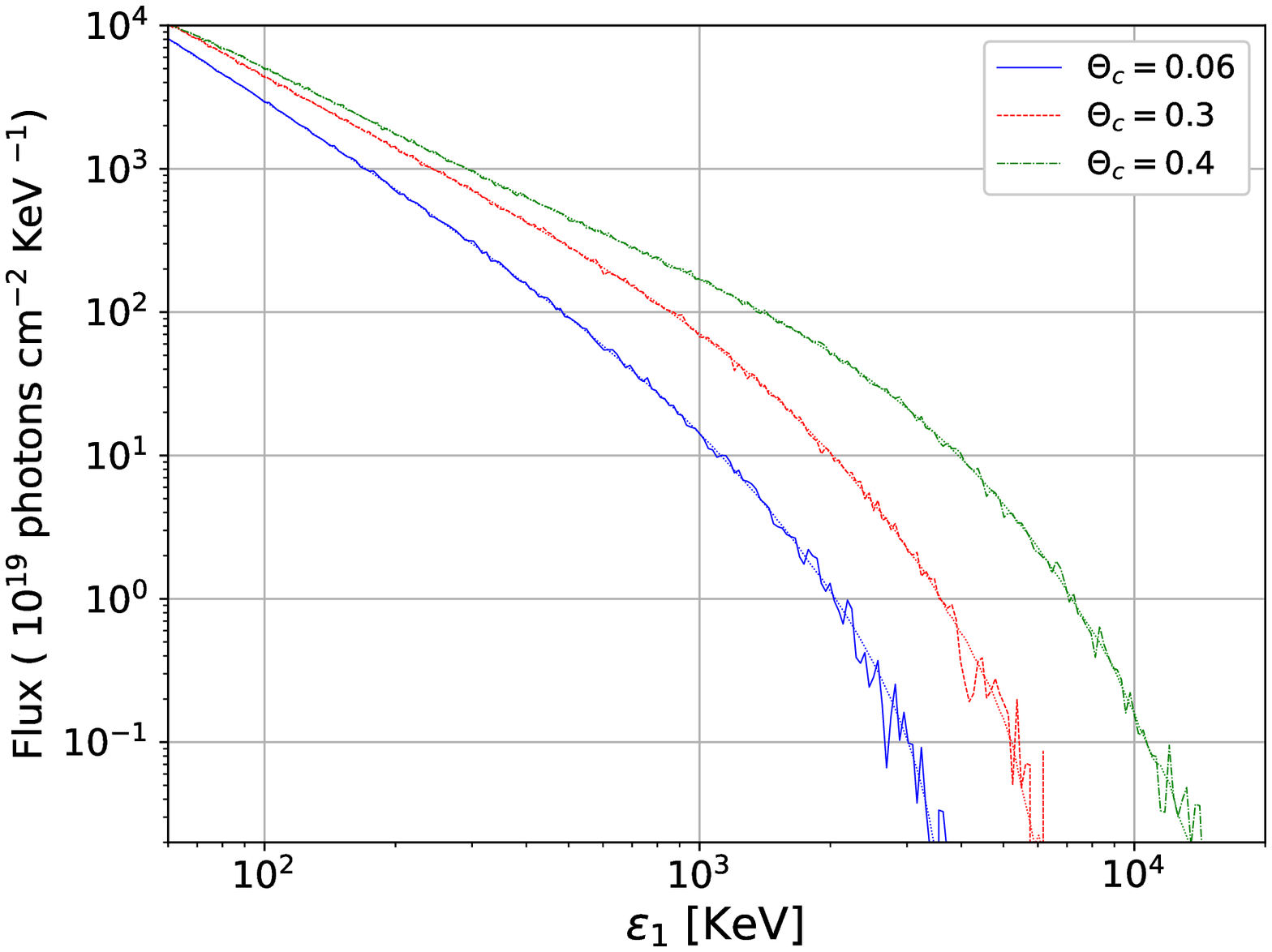}
  \includegraphics[width=7cm, angle=0]{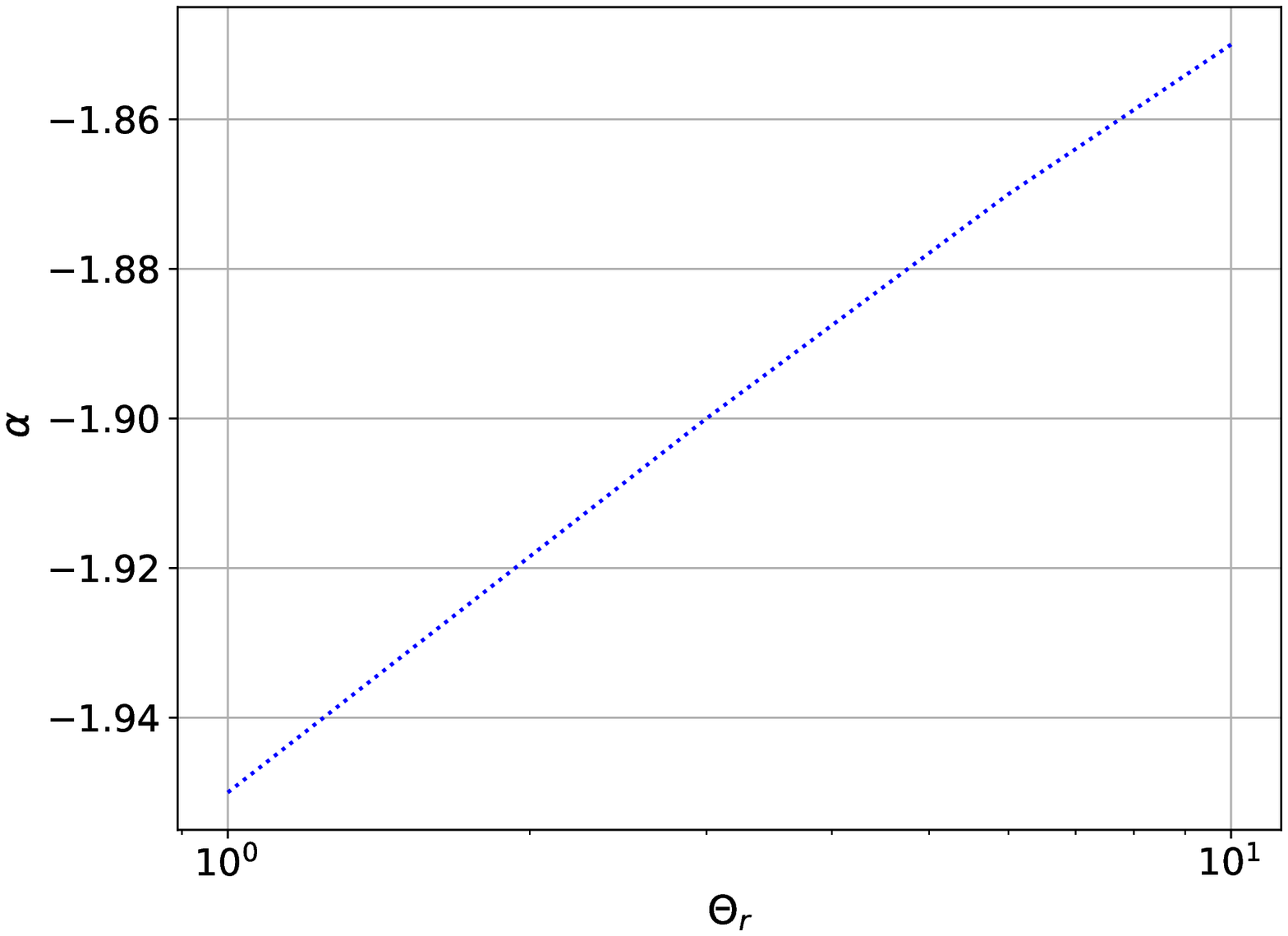}
    \includegraphics[width=7cm, angle=0]{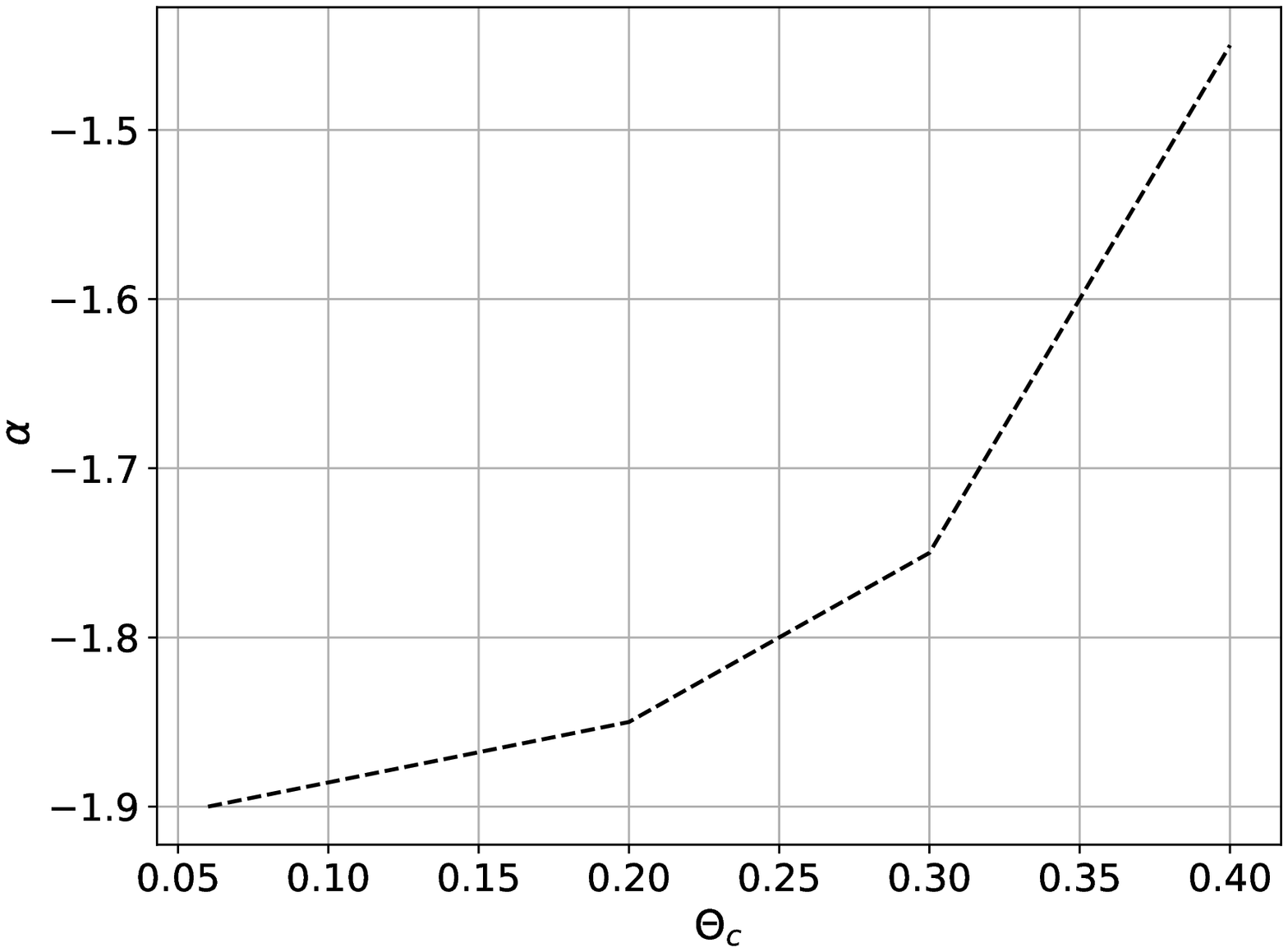}
    \includegraphics[width=7cm, angle=0]{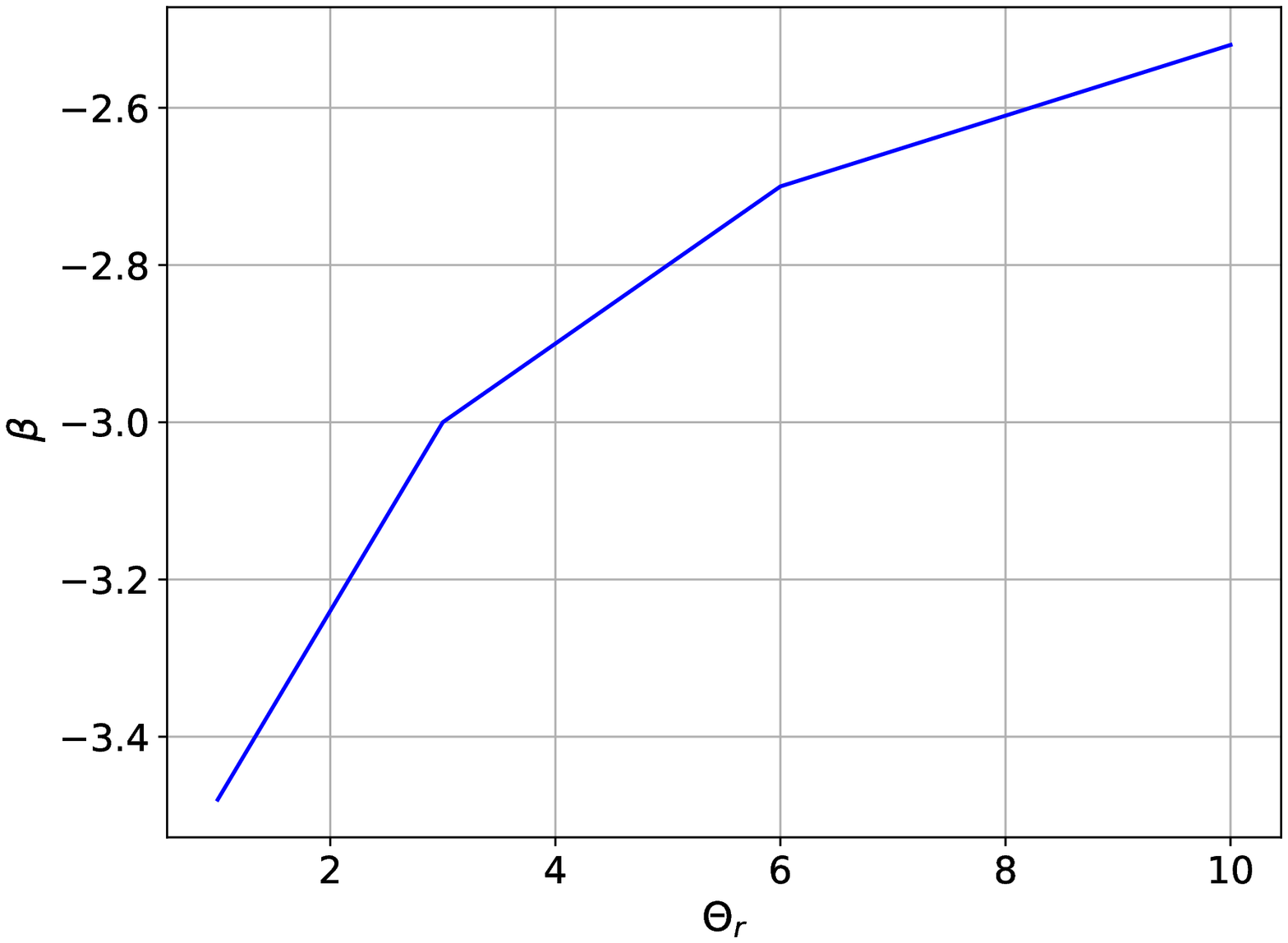}
        \includegraphics[width=7cm, angle=0]{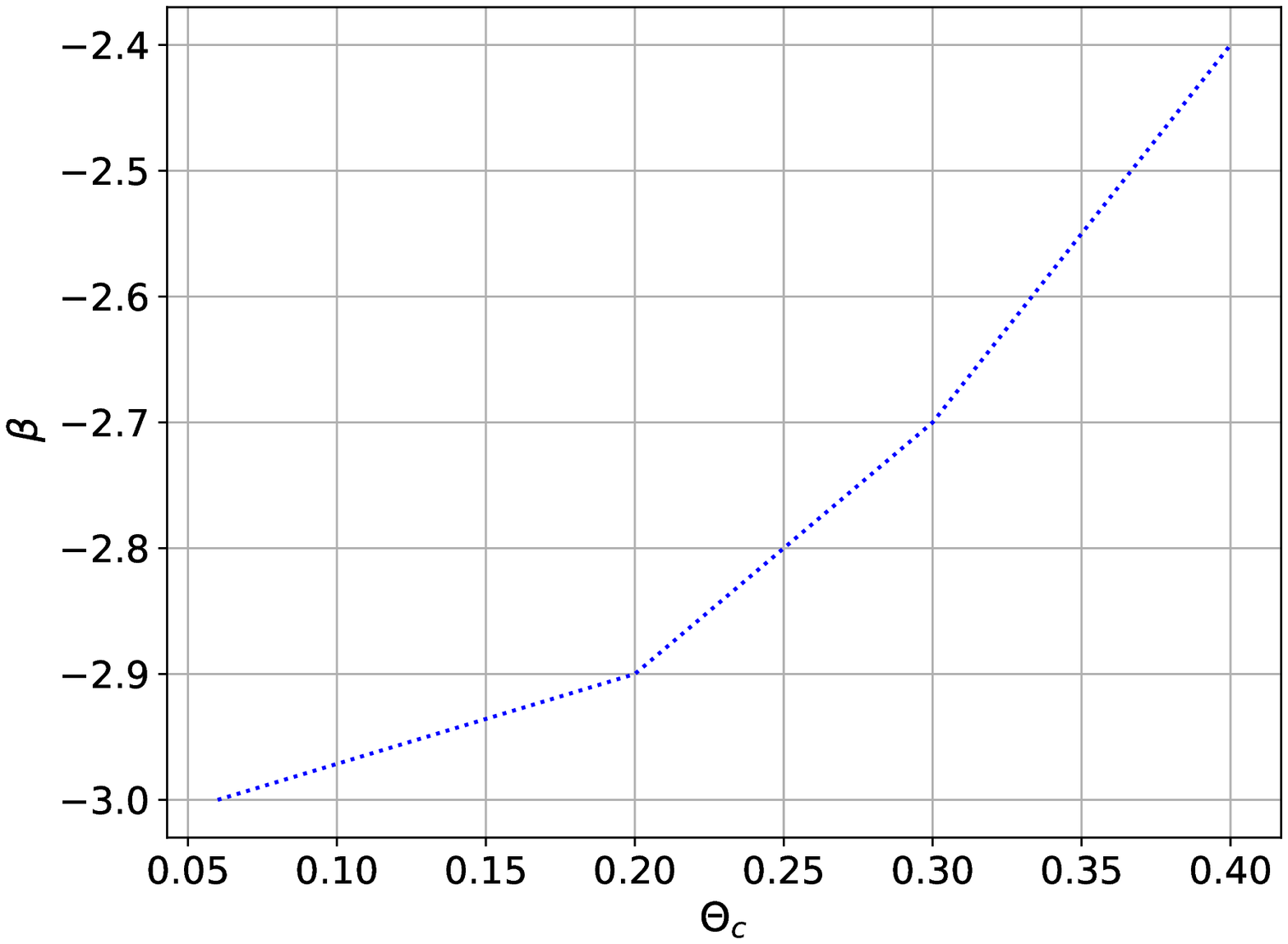}
    \includegraphics[width=7cm, angle=0]{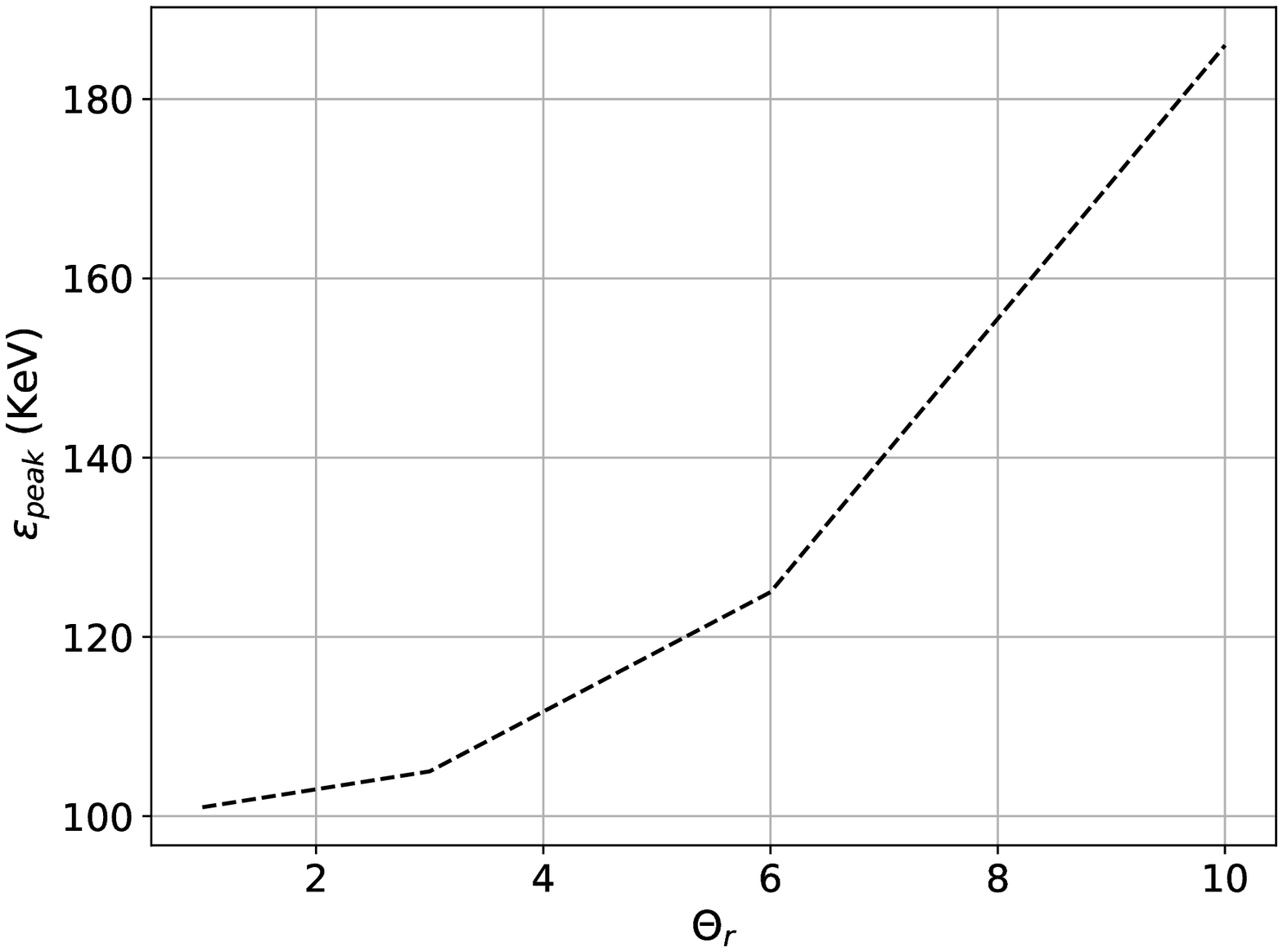}
            \includegraphics[width=7cm, angle=0]{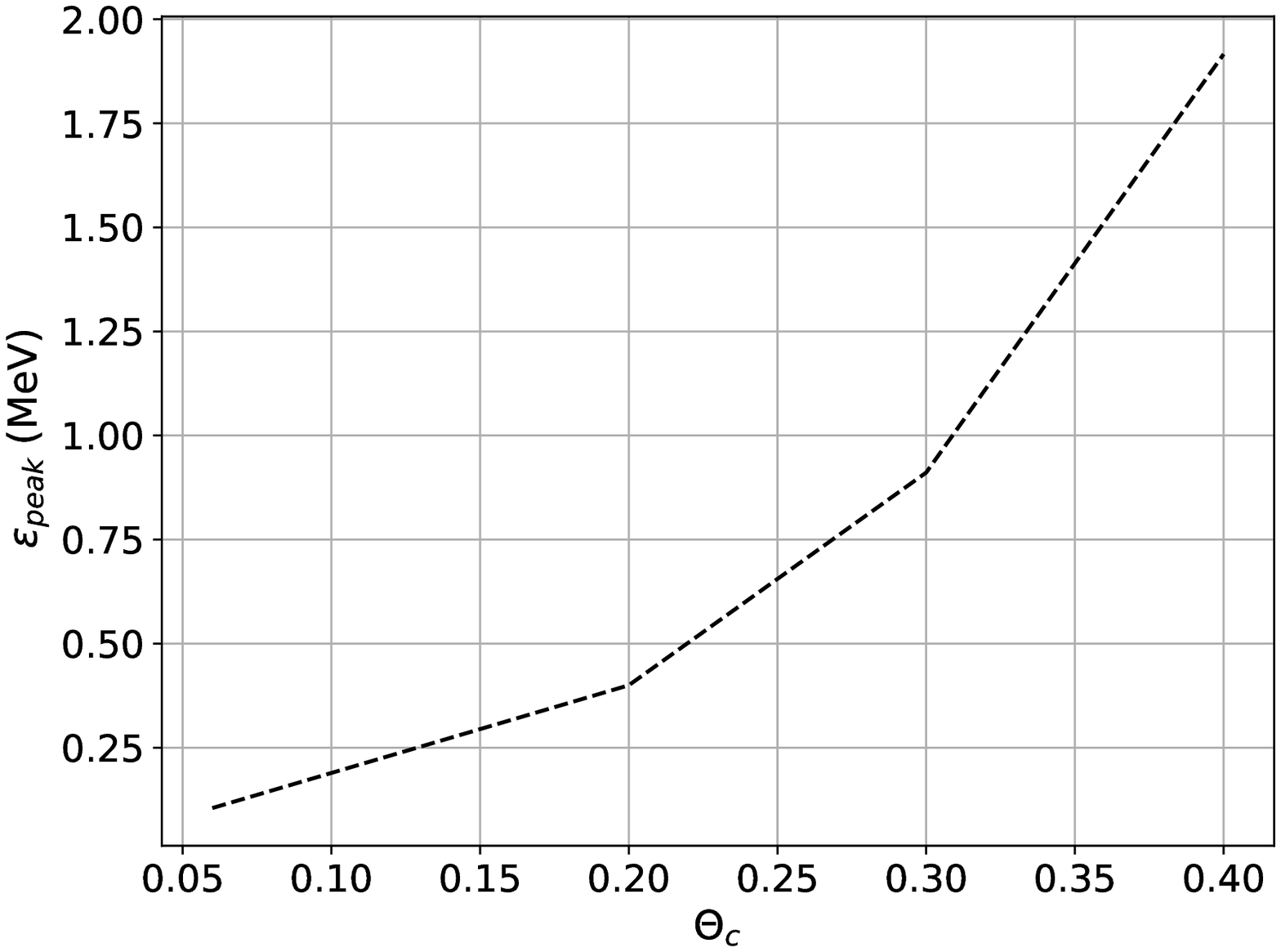}
\caption{\textbf{First row : Dependence of spectra on various choices of pair temperatures $\Theta_r=1,3,10$ choosing cork temperature $\Theta_c=0.06$ (left panel) and $\Theta_c=0.06,0.3,0.4$ keeping $\Theta_r=3$ (right panel). In both the panels, the overplotted dotted curves are corresponding smoothed spectra by applying Savitzky–Golay filter. Second row : variation of $\alpha$ with $\Theta_r$ (left) and $\Theta_c$ (left). Third row : $\beta$ as a function of $\Theta_r$ (left) and $\Theta_c$ (left). In the bottom row, dependence of $\varepsilon_{peak}$ is shown upon $\Theta_r$ (left panel) and $\Theta_c$ (right panel). For all the panels, $\theta_{obs}=0.105$ rad, $\theta_j=0.1$ rad and $\gamma=20$.}}
\label{lab_Spectrum_Tr_vary_multi}
 \end{center}
\end{figure*}

\textbf{The spectrum evolves with the seed photons' energy which is governed by pair temperature $\Theta_r$. Corresponding spectral variation is shown in the top left panel of Figure \ref{lab_Spectrum_Tr_vary_multi} for $\Theta_r=1,3$ and $10$. Here $\gamma=20$, $\Theta_c=0.06$ and $\theta_{obs}=0.105$ rad. The spectrum gets harder when the plasma assumes higher temperature and subsequently emits more photons at higher energy. 
In the subsequent panels below the spectra, we plot the variation of $\alpha$ $\beta$ and $\varepsilon_{peak}$ as functions of $\Theta_r$. As $\Theta_r$ varies in range 1-10, $\alpha$ mildly changes from -1.95 to -1.85, $\beta$ changes from -3.5 to -2.5 while the spectral peak energy $\varepsilon_{peak}$ evolves from 102 KeV to 186 KeV.}

\textbf{As the photons lose or gain energy by multiple scattering inside the cork, all the spectral parameters are sensitives to $\Theta_c$. In the top right panel of Figure \ref{lab_Spectrum_Tr_vary_multi}, we show the spectra for different choices of $\Theta_c$ in the range $0.06-0.4$. Here $\gamma=20$, $\Theta_r=3$ and $\theta_{obs}=0.105$ rad are kept constant. As the electrons are more energetic in the hotter cork, the spectra are harder showing decrease in magnitudes of $\alpha$, $\beta$ and increase in $\varepsilon_{peak}$, respectively shown in lower panels in the right column of Figure \ref{lab_Spectrum_Tr_vary_multi}. As $\Theta_c$ increases from $0.06$ to $0.4$, $\alpha$ varies from -1.9 to -1.55, $\beta$ changes from -3.0 to -2.4 and $\varepsilon_{peak}$ covers a large range between 187 KeV and 1.19 MeV. All these values are within the observed ranges seen in GRB prompt phase observations.}


\section{Summary}
\label{sec_summary}
In this letter we have considered {Comptonized }pair equilibrium {and bremsstrahlung} spectra near the centre of the star when  gamma ray burst takes place. These seed photons interact with a radially expanding stellar cork outside the stellar surface and are backscattered after undergoing Compton scattering with the relativistic electrons within the cork. The backscattered photons are then observed by an observer situated at angle $\theta_{obs}$ from the jet axis. 

The obtained spectra have a {negative} low energy photon index $\alpha$ {and} steeper high energy photon index $\beta$. 
\textbf{Our model predicts a large range of parameters $\alpha=-1.95$ to $-1.1$, $\beta=-3.5$ to $\sim -2.4$ and $\varepsilon_{peak}=$ a few KeV to few $10\times$ MeV. 
}
In the observed surveys, the peak values of low \textbf{and high } energy photon indices for the GRB population are obtained to be $\alpha=-1$  $\beta\sim -2.5$ and $\varepsilon_{peak} \sim 1$ MeV \citep{2000ApJS..126...19P,2006ApJS..166..298K,20115MNRAS.418L.109G, 2015AdAst2015E..22P} \textbf{with a large of variation their respective ranges in the GRB population}. 
\textbf{The obtained range of all the spectral parameters that govern the GRB prompt phase spectra are consistent with the observed ranges.} 
Hence, the modification makes the spectra in the backscattering model to be consistent with observations keeping all other findings in VPE21 unchanged.
In future works, we will shed light on the analytic understanding of the high energy photon indices $\beta$ and their dependence on physical parameters of the system. The evolution of pair plasma by expansion and subsequent emission of low energy seed photons can potentially contribute to the afterglow that we aim at examining further.

\acknowledgments
AP wishes to acknowledge support from the EU via ERC consolidator grant $773062$ (O.M.J.). MKV acknowledges the PBC program from the government of Israel {and Hüsne Dereli Bégué, Damien Bégué and Christoffer Lundman}  for important discussions. DE acknowledges support from The Israel Science Foundation grant 2131.

\bibliography{ref1}{}
\bibliographystyle{aasjournal}

\end{document}